\documentclass[sigconf]{acmart}

\def\BibTeX{{\rm B\kern-.05em{\sc i\kern-.025em b}\kern-.08emT\kern-.1667em\lower.7ex\hbox{E}\kern-.125emX}}

\usepackage[utf8]{inputenc}
\usepackage{graphicx}
\usepackage{amsmath}
\usepackage{algorithm}
\usepackage{algorithmic}
\usepackage{textcomp}
\usepackage{CJKutf8}
\usepackage{subcaption}
\usepackage{balance}

\newcommand{\sheader}[1]{{\flushleft \textit{#1}}}

\newcommand{\eg}{\textit{e.g.}}

\newcommand{\ie}{\textit{i.e.}}

\newcommand{\B}{\mathcal{B}}
\renewcommand{\L}{\mathcal{L}}

\newcommand{\N}{\mathcal{N}}
\newcommand{\R}{\mathbb{R}}

\newcommand{\dpsr}{DPSR}
\newcommand{\DPSR}{DPSR}

\newcommand\blfootnote[1]{%
  \begingroup
  \renewcommand\thefootnote{}\footnote{#1}%
  \addtocounter{footnote}{-1}%
  \endgroup
}

\renewcommand\footnotetextcopyrightpermission[1]{} % removes footnote with conference information in first column

\begin{document}

\clearpage\pagestyle{empty}

\title[Towards Personalized and Semantic Retrieval via Embedding Learning]{Towards Personalized and Semantic Retrieval:  An End-to-End Solution for E-commerce Search via Embedding Learning}

\author{Han Zhang$\,^{1\dagger}$, Songlin Wang$\,^{1\dagger}$, Kang Zhang$\,^{1}$, Zhiling Tang$\,^{1}$, Yunjiang Jiang$\,^{2}$, Yun Xiao$\,^{2}$, Weipeng Yan$\,^{1,2}$, Wen-Yun Yang$\,^{2*}$}
\affiliation{%
  \institution{$^{1}\,$JD.com, Beijing, China}
  \institution{$^{2}\,$JD.com Silicon Valley Research Center, Mountain View, CA, United States}
  \{zhanghan33, wangsonglin3, zhangkang1, tangzhiling, yunjiang.jiang, xiaoyun1, paul.yan, wenyun.yang\}@jd.com
}

\renewcommand{\shortauthors}{Han Zhang, et al.}


\begin{abstract}
Nowadays e-commerce search has become an integral part of many people's shopping routines.
Two critical challenges stay in today's e-commerce search: how to retrieve items that are semantically relevant but not exact matching to query terms, and how to retrieve items that are more personalized to different users for the same search query. In this paper, we present a novel approach called DPSR, which stands for Deep Personalized and Semantic Retrieval, to tackle this problem. Explicitly, we share our design decisions on how to architect a retrieval system so as to serve industry-scale traffic efficiently and how to train a model so as to learn query and item semantics accurately. Based on offline evaluations and online A/B test with live traffics, we show that DPSR model outperforms existing models, and DPSR system can retrieve more personalized and semantically relevant items to significantly improve users' search experience by +1.29\% conversion rate, especially for long tail queries by +10.03\%. As a result, our DPSR system has been successfully deployed into JD.com's search production since 2019.
\blfootnote{$^\dagger\,$ Both authors contributed equally}
\blfootnote{$^*\,$ Corresponding author}
\end{abstract}

\maketitle

\section{Introduction}
\begin{figure}
    \centering
    \includegraphics[width=40mm]{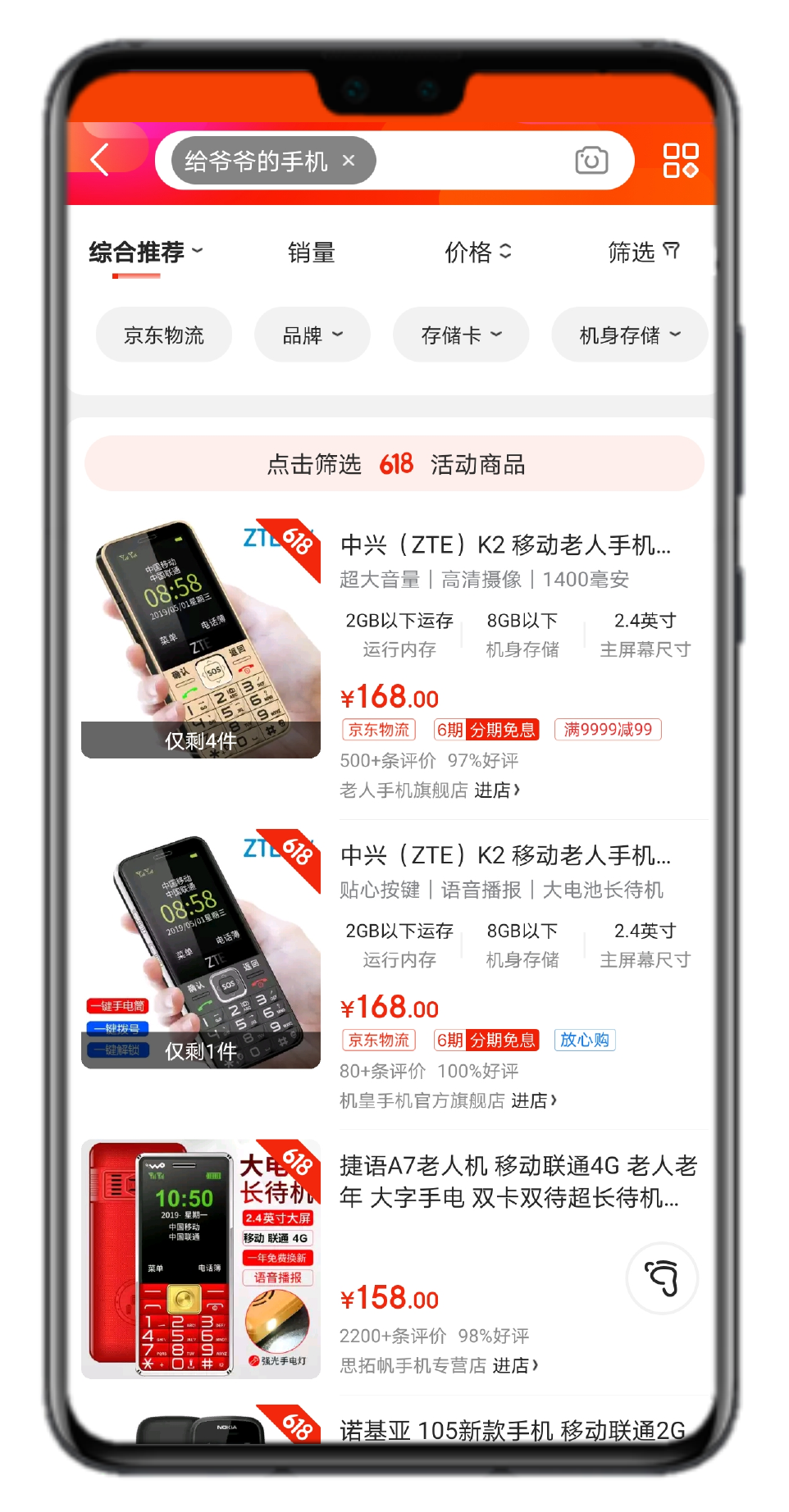}
    \vspace{-2mm}
    \caption{Search interface on JD's e-commerce mobile app. 
    %Here our {\dpsr} system helps search query (``cellphone for grandpa'' in Chinese) retrieve semantically related items with titles containing ``senior mobile phone'' in Chinese.
    }
    \vspace{-5mm}
    \label{fig:screenshot}
\end{figure}

Over the recent decades, online shopping platforms (e.g., Ebay, Walmart, Amazon, Tmall, Taobao and JD) have become increasingly popular in people's daily life. E-commerce search, which helps users to find what they need from billions of products, is an essential part of those platforms, contributing to the largest percentage of transactions among all channels~\cite{Liu:2017:CRO:3097983.3098011, Sondhi:2018:TQE:3209978.3210152, Sorokina:2016:ASJ:2911451.2926725}. 
For instance, the top e-commerce platforms in China, \eg, Tmall, Taobao and JD, serve hundreds of million active users with gross merchandise volume of hundreds of billion US dollar. 
In this paper, we will focus on the immense impact that deep learning has recently had on the e-commerce search system. At a glance, Figure~\ref{fig:screenshot} illustrates the user interface for searching on JD's mobile app.

\subsection{Three Components of Search System}
\begin{figure}
    \centering
    \includegraphics [width=0.48\textwidth]{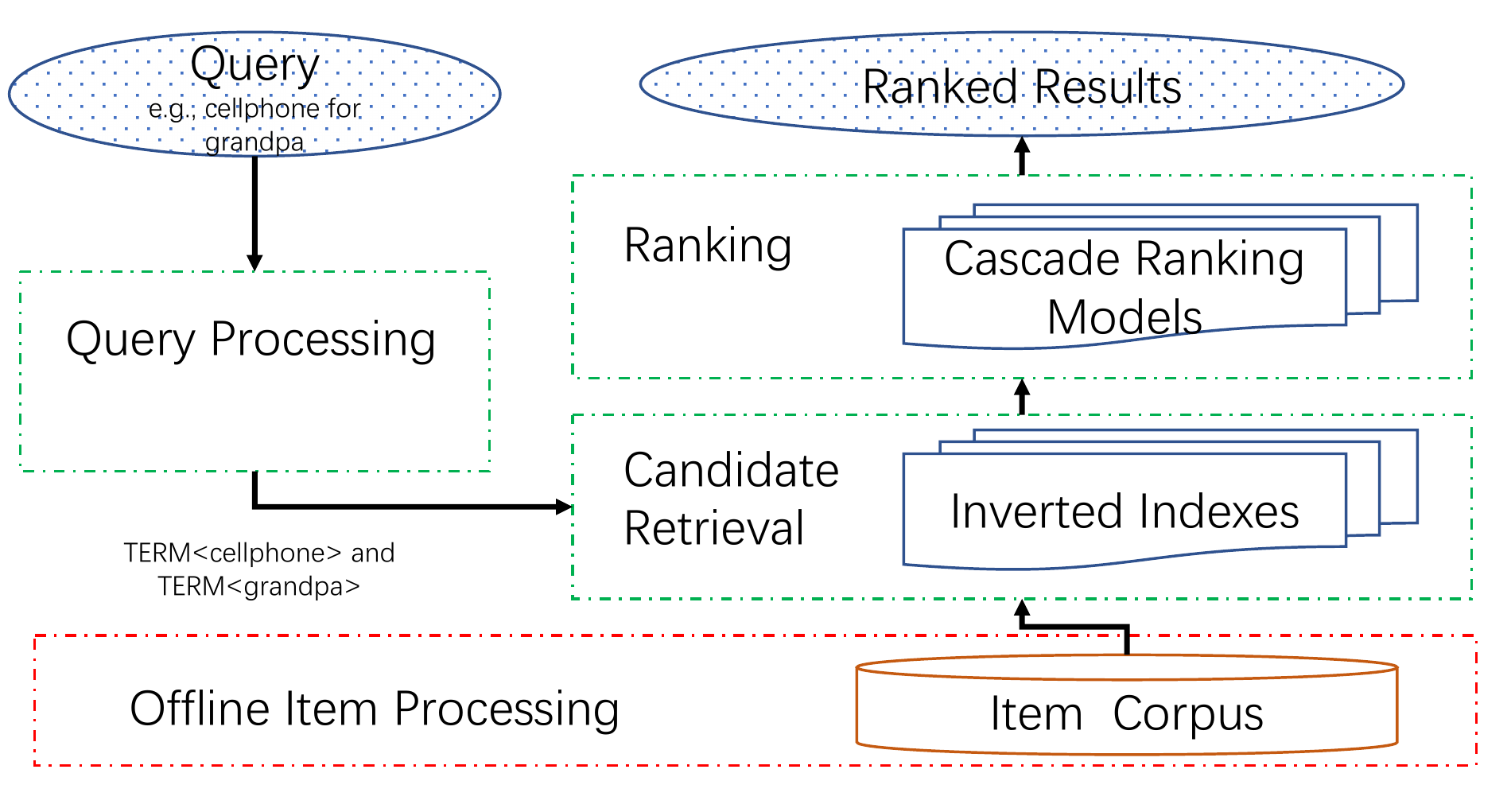}
    \vspace{-3mm}
    \caption{Major stages of an e-commerce search system.}
    \label{fig:search-engine}
    
\end{figure}

Figure~\ref{fig:search-engine} illustrates a typical e-commerce search system with three components, query processing, candidate retrieval, and ranking.

\emph{Query Processing} rewrites a query (\eg, ``cellphone for grandpa'') into a term based presentation (\eg,  [TERM cellphone] AND [TERM grandpa]) that can be processed by downstream components. This stage typically includes tokenization, spelling correction, query expansion and rewriting.

\emph{Candidate Retrieval} uses offline built inverted indexes, to efficiently retrieve candidate items based on term matching. This step greatly reduces the number of items from billions to hundreds of thousands, in order to make the fine ranking feasible.

\emph{Ranking} orders the retrieved candidates based on factors, such as relevance, predicted conversion ratio, etc. A production system may have cascading ranking steps, which sequentially apply simpler to more complex ranking functions from upstream to downstream.

In this paper, we focus solely on the \emph{candidate retrieval} stage to achieve more personalized and semantic search results, since this stage contributes the most bad cases in our search production.
Based on our analysis, around $20\%$ dissatisfaction cases of search traffic of JD.com, one of the largest e-commerce search engine in the world, can be attributed to the failure of this stage.
How to deal with that in the ranking stage is out of scope for this paper, but will be our future work.

% How to deal with that in the ranking stage is out of scope for this paper, but will be our future work.
% We note that while many deep neural network models, such as DSSM~\cite{Huang:2013:LDS:2541176.2505665}, Relevance-based Word Embedding Model~\cite{Zamani:2017:RWE:3077136.3080831}, and Neural IR Models~\cite{Hui:2018:CCN:3159652.3159689} have been proposed, they mainly focus on how to measure relevance between a query and a few hundreds of items in the ranking stage. 
% Our work in this paper solves a very different problem, namely how to efficiently retrieve personalized and semantically relevant items first in the retrieval stage out of hundreds of millions of candidates. We are \emph{not} dealing at all with the ranking stage.
%A production search system usually requires retrieving tens of thousands of relevant items from billions of items within tens of milliseconds.

\subsection{Two Challenges in Candidate Retrieval}
How to efficiently retrieve more personalized and semantically relevant items remains two major challenges in modern e-commerce search engines.

\emph{Semantic Retrieval Problem} refers to that,
items that are semantically relevant but do not contain the exact terms of a query cannot be retrieved by traditional inverted indexes.
As reported in~\cite{Li:2014:SMS:2692909.2692910}, the most critical challenge for search systems is term mismatch between queries and items, especially for e-commerce search, where item titles are often short. 
Traditional web search often uses query rewriting to tackle this problem, which transforms the original query to another similar query that might better represent the search need. However, it is hard to
ensure that the same search intention can be kept through a ``middle man'', \ie, rewritten queries, and there is also no guarantee that relevant items containing different terms can be retrieved via a limited set of rewritten queries. 

\emph{Personalized Retrieval Problem} refers to that,
traditional inverted indexes cannot retrieve different items according to the current user's characteristics, \eg, gender, purchase power and so on.
For example, we would like to retrieve more women's T-shirt if the user is female, and vice versa. Some rule-based solutions have been used in our system for years include that, 1) indexing tags for items, \eg, purchase power, gender and so on, the same way as tokens into the inverted index, 2) building separate indexes for different group of users. However, these previous approaches are too hand-crafted. Thus, they are hard to meet more subtle personalization needs.

\subsection{Our Contributions}
In this paper, we propose {\dpsr}: Deep Personalized and Semantic Retrieval,
to tackle the above two challenges in a leading industrial-scale e-commerce search engine. The contributions of our work can be summarized as follows.

In Section~\ref{sec:overview}, we present an overview of our full {\dpsr} embedding retrieval system composed of offline model training, offline indexing and online serving. We share our critical design decisions for productionizing this neural network based candidate retrieval into an industry-level e-commerce search engine.

% In Section~\ref{sec:overview}, we introduce our full {\dpsr} embedding retrieval system, which uses a neural network model to learn embedding for items and queries based on their semantics, and an embedding retrieval component to find nearest neighbor items for a query based on their embeddings. We share our critical design decisions for productionizing neural network based candidate retrieval into an industry-level e-commerce search engine.

In Section~\ref{sec:model}, we develop a novel neural network model with a two tower architecture, a multi-head design of query tower, an attention based loss function, a negative sampling approach, an efficient training algorithm, and human supervision data, all of which are indispensable to train our best performing models.

In Section~\ref{sec:system}, we present our efforts on building a large-scale deep retrieval training system where we significantly customize the off-the-shelf TensorFlow API for online/offline consistency, input data storage and scalable distributed training, and on building an industrial-scale online serving system for embedding retrieval.

In Section~\ref{sec:experiments}, we conduct extensive embedding visualization, offline evaluation and online AB test to show that our retrieval system can help to find semantically related items and significantly improve users' online search experience, especially for the long tail queries, which are difficult to handle in traditional search systems (\ie, improving conversion rate by around $10\%$).

\section{Related Work}
%We review recent progress in candidate retrieval, deep learning based methods for semantic relevance, and embedding retrieval methods.

\subsection{Traditional Candidate Retrieval}
For candidate retrieval, most research focuses on learning query rewrites~\cite{Guo:2008:UDM:1390334.1390400,Bai:2018:SQN:3219819.3219897} as an indirect approach to bridge vocabulary gap between queries and documents.
Only a few new approaches, including latent semantic indexing (LSI)~\cite{Deerwester90indexingby} with matrix factorization, probabilistic latent semantic indexing (PLSI)~\cite{Hofmann:1999:PLS:312624.312649} with probabilistic models, and semantic hashing~\cite{Salakhutdinov:2009:SH:1558385.1558446} with an auto-encoding model, have been proposed. 
All of these models are unsupervised learned from word co-occurrence in documents, without any supervised labels.
Our approach differs from the previous methods in that we train a supervised model to directly optimize relevance metrics based on a large-scale data set with relevant signals, \ie, clicks.

% All of these studies inseminate the idea of learning a dense representation for queries and documents to enable semantic retrieval. However, all of them are unsupervised learned based on co-occurrence of words in documents. We propose a customized two-tower model to train on query-item paired data (clicks), which can better model what is important for determining relevance, and more importantly, we share important system designs and model training practices for enabling semantic retrieval in a production system, which serves millions of users. 

%(As reported by a recent survey~\cite{introduction-neural-information-retrieval}, most of recent IR research focuses on relevance scoring in the ranking stage)
%Deep N Semantic Hashing (seminal work), unsupervised document encoding,  

\begin{figure*}[t]
    \centering
    \includegraphics[width=\textwidth]{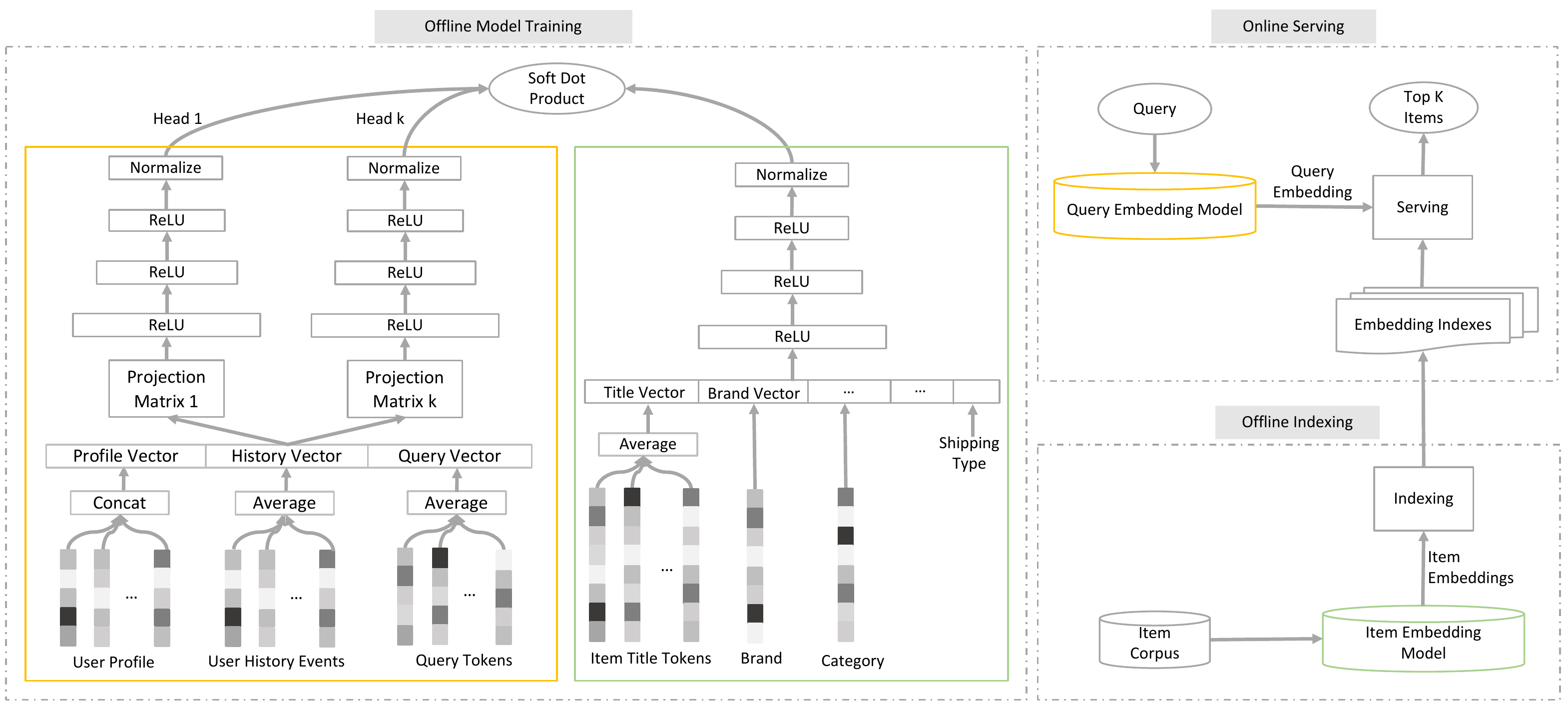}
    \caption{Overview of our {\dpsr} retrieval system.}
    \label{fig:overview}
\end{figure*}

\subsection{Deep Learning Based Relevance Model}
With the success of deep learning, a large number of neural network based models have been proposed to advance traditional information retrieval (IR) methods (\eg, BM2.5) and learning to rank methods~\cite{l2rSurvey} in the manner of learning semantic relevance between queries and documents. See~\cite{Li:2014:SMS:2692909.2692910} and ~\cite{introduction-neural-information-retrieval} for a comprehensive survey in semantic match and deep neutral network based IR. Particularly, DSSM~\cite{Huang:2013:LDS:2541176.2505665} and its following work  CDSSM~\cite{Shen:2014:LSR:2567948.2577348} have pioneered the work of using deep neural networks for relevance scoring. 
%DSSM consists of two fully-connected towers for embedding queries and documents and optimizes a softmax loss based on the cosine distance of the learned embeddings. 
%As these methods purely focus on semantic representations of queries and documents, 
Recently, new models including DRMM~\cite{Guo:2016:DRM:2983323.2983769}, Duet~\cite{Mitra:2017:LMU:3038912.3052579} have been further developed to include traditional IR lexical matching signals (\eg, query terms importance, exact matching) in neural networks. However, as reported by~\cite{introduction-neural-information-retrieval}, most of the proposed works in this direction focus on ranking stage, where the optimization objectives and requirements are very different from candidate retrieval that our work in this paper focuses on.

Two tower architecture for deep neural network has been widely adopted in existing recommendation works~\cite{yang2020mixed, yi2019sampling} to further incorporate item features. This model architecture is also known as dual encoder in natural language processing~\cite{cer2018universal,henderson2017efficient}. Here we propose a more advanced two tower model which is composed of a multi-head tower for query and an attention loss based on soft dot product instead of simple inner product.

% Thus they are not applicable
% limited to re-ranking top-n document based on relevance scores, and are not designed for candidate retrieval that we are tackling in this paper.

% We note that, while our DR model is similar to DSSM as we will discuss in Section~\ref{sec:model}, and our experiment in Section~\ref{sec:experiments}, our model is specifically designed for task (in terms of loss function and training method), which performs much better than DSSM in our task. 

\subsection{Embedding Retrieval in Search Engine}
Recently, embedding retrieval technologies have been widely adopted in modern recommendation and advertising systems~\cite{youtube2016, mind2019, tree2018}, while have not been widely used in search engine yet. We find a few works about retrieval problems in search engine~\cite{palangi2016deep,vu2017search}, while they have not been applied to industrial production system. To the best of our knowledge, we are one of the first practical explorations in this direction of applying embedding retrieval in industrial search engine system.
\section{Overview of Embedding Retrieval System}
\label{sec:overview}

%Conceptually, our proposed embedding retrieval approach is based on the following intuition: we can first train a deep model, which is able to project queries and items to embeddings based on their semantics, and then build a system to efficiently retrieve relevant items for a query in the embedding space. 

Before we present the details, let us first show a full picture of our embedding retrieval system.
Figure~\ref{fig:overview} illustrates our production system with three major modules as follows.

\sheader{Offline Model Training} module trains a two tower model consisting of a query embedding model (\ie, query tower) and an item embedding model (\ie, item tower) for the uses in online serving and offline indexing respectively. This two tower model structure is a careful and essential design to enable fast online embedding retrieval, which we will discuss more in Section~\ref{sec:model}. Moreover, We will also talk about our effort of optimizing offline training system in Section~\ref{sec:training_system}.

\sheader{Offline Indexing} module loads the item embedding model (\ie, the item tower) to compute all the item embeddings from the item corpus, and then builds an embedding index offline to support efficient online embedding retrieval.
As it is infeasible to exhaustively search over the item corpus of billions of items, to find similar item embeddings for a query embedding, we employ one of state-of-the-art algorithms~\cite{JDH17} for efficient nearest neighbor search of dense vectors. 

%The general ideas of the algorithm can be summarized as clustering the vectors by $k$-means, compressing the vectors by product quantization, and building an index based on the compressed vectors to reduce the computation to calculate distance between vectors. 

\sheader{Online Serving} module loads the query embedding model (\ie, the query tower) to transform any user input query text to query embedding, which is then fed to the item embedding index to retrieve $K$ similar items. Note that this online serving system has to be built with low latency of tens of milliseconds. Also, it must be scalable to hundreds of thousands queries per second (QPS), and flexible for agile iterations of experiments. We will talk about our efforts of building such an online serving system in Section~\ref{sec:serving_system}.

\section{Embedding Learning Model}
\label{sec:model}
In this section, we introduce the embedding learning model in a stepwise fashion, in the order of two tower architecture, multi-head design of query tower, attentive loss function, hybrid negative sampling, and human supervision data, all of which are indispensable to train our best performing model.

\subsection{Two Tower Model Architecture} 
\label{sec:two_tower_model}
As shown in offline model training module in Figure~\ref{fig:overview}, the model is composed of a query tower $Q$ and an item tower $S$. For a given query $q$ and an item $s$, the scoring output of the model is
\begin{equation*}
f(q, s) = G(Q(q), S(s))
\end{equation*}
where $Q(q) \in \R^{d \times m}$ denotes query tower outputs of $m$ query embeddings in $d$-dimensional space. Similarly, $S(s) \in \R^{d \times n}$ denotes item tower outputs. The scoring function $G(.,.)$ computes the final score between the query and item. 
Researchers and practitioners often let query tower $Q$ and item tower $S$ both output one single embedding, \ie, $m=1$ and $n=1$, and choose $G$ as inner product, \ie,  $G(Q(q), S(s)) = Q(q)^\top S(s)$ where the superscript $^\top$ denotes matrix transpose. This simplest setup has been proved to be successful in many applications~\cite{youtube2016}.

The key design principle for such two tower architecture is to make the query embedding and the item embedding independent on each other after the model is trained. So we can compute them separately. All item embeddings can be computed offline in order to build an item embedding index for fast nearest neighbor search online, and the query embedding can be computed online to handle all possible user queries. 
Even though the embeddings are computed separately, due to the simple dot product interaction between query and item towers, the query and item embeddings are still theoretically in the same geometric space. Thus, finding $K$ nearest items for a given query embedding is equivalent to minimizing the loss for $K$ query item pairs where the query is given. 

In below sections, we will introduce a novel design of query tower $Q$ and an interaction function $G$ to achieve outperforming and explainable retrieval results. Since item representations are normally straightforward, we still keep the item tower $S$ typically simple. It concatenates all item features as input layer, then goes through multi-layer perceptron (MLP) of fully connected Rectified Linear Units (ReLU) to output a single item embedding, which is finally normalized to the same length as query embedding, as shown in the right side of offline model training panel in Figure~\ref{fig:overview}. Similar MLP structure can be found in previous work~\cite{youtube2016}.

\subsection{Query Tower with Multi-heads}
\label{sec:multi_head}
As shown in the left side of offline model training panel in Figure~\ref{fig:overview}, query tower differs from item tower in two places, 1) a projection layer that projects the one input dense representation to $K$ dense representations. Another choice here is to use $K$ independent embedding set, but it requires larger model size. In practice, we choose the projection layer to achieve similar results but with much smaller model size. 
2) $K$ separate encoding MLPs, each of which independently outputs one query embedding that potentially would capture different intention for the query. We refer to these $K$ output embeddings as \emph{multi-head representations}.

These multiple query embeddings provide rich representations for the query's intentions. Typically, we find in practice that it could capture different semantic meanings for a polysemous query (\eg, ``apple''), different popular brands for a product query (\eg, ``cellphone''), and different products for a brand query (\eg, ``Samsung''). %Moreover, we found in practice that using multiple embeddings also provide model ensemble effect, since we can think of the $K$ separate projections and encoding MLPs as $K$ separately trained models. 

It is worth mentioning that the encoding layer can use any other more powerful neural network, such as Recurrent Neural Network (RNN) and other state-of-the-art transformer based models~\cite{Vaswani_nips2017,bert,gpt2}.
In a separate offline study, we have achieved similar or slightly better results with these advanced models.
However, we would like to emphasize that a simple MLP is more applicable to our industrial production modeling system, since it is much more efficient for both offline training and online serving, which means that we are able to feed more data to the model training, and deploy fewer machines to serve the model.
These are strong deal breakers in industrial world.

\subsection{Optimization with Attention Loss}
\label{sec:model_training}
Apart from the single embedding and inner product setup, here we develop a more general form for multiple query embeddings. 
As a shorthand, we denote each output of query tower $Q(q)$ as $\{e_{1}, e_{2}, \ldots, e_{m}\}$ where $e_{i} \in \R^{d}$, and the single output of item tower $S(s)$ as $g \in \R^{d}$. Then the soft dot product interaction between query and item can be defined as follows,
\begin{equation}
G(Q(q), S(s)) = \sum_{i=1}^{m} w_i e_{i}^\top g.
\label{eq:soft_dot}
\end{equation}
This scoring function is basically a weighted sum of all inner products between $m$ query embeddings and one item embedding. The weights $w_i$ are calculated from softmax of the same set of inner products,
\begin{equation*}
w_i = \frac{\exp(e_{i}^\top g / \beta)}{\sum_{j=1}^{m} \exp(e_{j}^\top g / \beta)},
\end{equation*}
where $\beta$ is the temperature parameter of softmax. Note that the higher the $\beta$ is, the more uniform the attention weights appear. If $\beta \rightarrow 0$, then the soft dot product in Equation~(\ref{eq:soft_dot}) would be equivalent to selecting the largest inner product, \ie, $\max_i e_{i} \top g$.

A typical industrial click log data set usually contains only click pairs of query and item. The pairs are usually relevant, thus can be treated as positive training examples. Besides that, we also need to collect negative examples by various sampling techniques that we will talk about later in Section~\ref{sec:batch_negative}. 
Let us define the set $\mathcal{D}$ of all training examples as follows, 
\begin{equation}
\mathcal{D} = \left\{\left(q_i, s^+_i, \mathcal{N}_i\right) \; \left| \; i,\; r(q_i, s_i^+) = 1, \; r(q_i, s^-_j) = 0\; \forall\; s^-_j \in \mathcal{N}_i \right\}\right.,
\label{eq:dataset}
\end{equation}
where each training example is a triplet composed of, a query $q_i$, a positive item $s_i^+$ that is relevant to the query denoted as $r(q_i, s_i^+) = 1$, and an negative item set $\mathcal{N}_i$ where every element $s^-_j$ is irrelevant to the query, denoted as $r(q_i, s^-_j) = 0$.
Then we can employ hinge loss with margin $\delta$ over the training data set $\mathcal{D}$ as follows,
\begin{equation*}
\small\mathcal{L}(\mathcal{D}) = \sum_{(q_i, s^+_i, \mathcal{N}_i) \in \mathcal{D}} \sum_{s^-_j \in \mathcal{N}_i}
\max\left(0,\delta - f(q_i, s^+_i) + f(q_i, s^-_j)\right).
\label{eq:loss}
\end{equation*}

The above attention loss is only applied in the offline training. During the online retrieval, each query head retrieves the same number of items. Then all the items will be sorted and cut off based on their inner products with one of the heads.

\subsection{Click Logs with Negative Sampling}
\label{sec:batch_negative}

Training a deep model requires a huge amount of data.  We explore click logs, which represents users' implicit relevance feedback and consists of a list of queries and their clicked items, to train our embedding retrieval model. Intuitively, we can assume that an item is relevant, at least partially, to the query if it is clicked for that query. Formally, we can consider click logs as a special case of data set with only positive examples. Then how to efficiently collect negative examples is a crucial question here. In our practice, we employ a hybrid approach that mixes two sources of negative samples, including random negatives and batch negatives. 

\subsubsection{Random Negatives}
Random negative set $\N_i^{rand}$ are uniformly sampled from all candidate items. Formally, given a set of all $N$ available items, we draw a random integer variable from a uniform distribution $i \sim Uniform(1, N)$, and take the $i$-th element from the item set into random negative set $\N_i^{rand}$.
However, if we apply this uniform sampling in a straightforward way, it would be very computational expensive, since each negative sample has to go through the item tower, not to mention the cost for sampling those negative examples and fetching their features. To minimize the computational cost while retaining its effect, we use the same random negative set for all training examples in a batch. In practice, we found the results are similar to that using pure random negatives but the training speed is much faster.

\subsubsection{Batch Negatives}
Batch negative set $\N_i^{batch}$ are collected by permuting the positive query item pairs in a training batch. In detail, for a training batch
\[
\B = \{(q_i, s_i^+, \N_i)\; |\; i\;\},
\]
we can collect more negative examples for the $i$-th example as 
\[
\N_i^{batch} = \left\{s^+_k \;|\; k \neq i, 1\le k \le |\B| \right\}.
\] 
We can see that batch negatives are basically sampled according to item frequency in the dataset. These randomly generated query and item pairs are very unlikely to be relevant by chance. Specifically, the chance is equal to that two randomly drawn click logs having relevant items for each other. Given a dataset of hundreds of millions of click logs, this chance is basically ignorable in terms of training accuracy. Also, the main advantage of the above batch negatives is the reuse of the item embedding computations. Each item embedding in the batch serves once as positive example, and $|\mathcal{B}| - 1$ times as negative examples for other queries in the batch, but with only one feature fetching and forward pass of the item tower.

\subsubsection{Mixing Ratio}
\label{sec:mixing_ratio}
Eventually, the complete negative item set $\N_i$ in Equation~(\ref{eq:dataset}) is a union set of above two sets, 
\[
\N_i = \N_i^{rand} \cup \N_i^{batch}.
\]
In our practice of e-commerce search retrieval, we find it is typically useful to have a mixing ratio parameter $0 \le \alpha \le 1$ for the composition of negative sample set. Formally, we use proportion $\alpha$ of random negatives, and proportion $(1-\alpha)$ of batch negatives. We find the value of $\alpha$ highly correlates with the popularity of items retrieved from the model (see Experiments), thus highly influential to online metrics. 
Intuitively, we can see that the mixing ratio $\alpha$ determines the item distribution in negative examples, from uniform distribution ($\alpha = 1$) to actual item frequency ($\alpha = 0$). In this manner, the model tends to retrieve more popular items for larger $\alpha$, as popular items appear relatively less frequently in negative examples.

\subsubsection{Summary}
% It is worth mentioning that unclicked items or skipped items in the same session are used as negative examples in DSSM~\cite{Huang:2013:LDS:2541176.2505665} and other click-based ranking models. However, an unclicked/skipped item in a session does not necessarily mean irrelevance, especially in e-commerce search where click rate is also impacted by price and other factors. Thus, though unclicked/skipped items might be a good choice for ranking models to distinguish between relevant and less relevant items in returned candidates, it is far from applicable for our retrieval model to find relevant candidates from all items (see Table~\ref{tab:batch_negative}).

We summarize the full training algorithm with batch negatives and random negatives in Algorithm~\ref{alg:batch_neg}. The computational complexity for each training step is $O(b^2)$, \ie, quadratic with the batch size $b$, since the batch negatives require an inner product between every query and item embedding pair in the batch. In practice, since the batch size is usually small, \eg, 64 or 128, the quadratic effect is actually much smaller than other computational cost, \ie, feature fetching, gradient computation, and so on. In fact, with batch negatives, the total convergence is actually faster, due to the efficient use of every item tower outputs.

\begin{algorithm}
\begin{algorithmic}[1]
\STATE \textbf{input}: Dataset $\mathcal{D}$, batch size $b$, max number of steps $T$, mixing ratio $\alpha$.
\FOR{$t = 1 \ldots T$} 
\STATE Sample a batch of $b$ examples $\B \subseteq \mathcal{D}^+$.
\STATE Sample a set of random negatives $\N^{rand}$ for this batch. Note that all examples in the batch shares this set.
\STATE Compute query head embeddings $Q(q)$ from query tower.
\STATE Compute item embeddings $S(s)$ for all item $s_i$ in the batch, and that in the random negative set $\N^{rand}$.
\STATE Compute loss function value $\L(\B)$ for this batch $\B$. The batch negatives $\N^{batch}$ are implicitly computed and included in the loss.
\STATE Update towers $Q$ and $S$ by back propagation.
\ENDFOR
\RETURN query tower $Q$ and item tower $S$.
\end{algorithmic}
\caption{{\DPSR} training algorithm}
\label{alg:batch_neg}
\end{algorithm}

\subsection{Human Supervision}
% user feedback, pairwise training
Beyond using click logs data, our model is also able to utilize additional human supervision to further correct corner cases, incorporate prior knowledge and improve its performance. The human supervision comes from three sources:

\begin{itemize}
\item \emph{Most skipped items} can be automatically collected from online logs~\cite{Joachims:2007:EAI:1229179.1229181}. These items and the associated queries can be used as negative examples.
\item \emph{Human generated data} can be collected based on domain knowledge as artificial negative query item pairs (\eg, cellphone cases are generated as negative items for query ``cellphone'', because they share similar product words literally but differ significantly in semantic meaning) and positive query item pairs (\eg, iPhone 11 items are generated as positive items for query ``newest large screen iphone'').
\item \emph{Human labels and bad case reports} are normally used to train relevance models~\cite{Yin:2016:RRY:2939672.2939677}. We also include them as both positive and negative examples in the training data set.
\end{itemize}
These human supervision data can be fed into the model as either positive query item pairs or an item in the random negative set.
\section{Embedding Retrieval System}
\label{sec:system}
We employ TensorFlow~\cite{tensorflow} as our training and online serving framework, since it has been widely used in both academia and industry. Particularly, it has the advantage of high-performance of training speed with static graph pre-built before training, and seamless integration between training and online serving. We built our system 
based on the high level TensorFlow API, called Estimator~\cite{estimator}. 
To ensure best performance and system consistency, we have also made significant efforts to abridge an off-the-shelf TensorFlow package and an industry level deep learning system.

\subsection{Training System Optimizations}
\label{sec:training_system}

\subsubsection{Consistency Between Online and Offline}
One of the common challenges for building a machine learning system is to guarantee the offline and online consistency. 
A typical inconsistency usually happens at the feature computation stage, especially if two separate programming scripts are used in offline data pre-processing and online serving system. In our system, the most vulnerable part is the text tokenization, carried on three times in data preprocessing, model training and online serving.
In aware of this, we implement one unique tokenizer in C++, and wrap it with a very thin Python SWIG interface~\cite{swig} for offline data vocabulary computation, and with TensorFlow C++ custom operator~\cite{tensorflow} for offline training as well as online serving. Consequentially, it is guaranteed that the same tokenizer code runs through raw data preprocessing, model training and online prediction.

\subsubsection{Compressed Input Data Format}
A typical data format for industrial search or recommendation training system is usually composed of three types of features, user features (\eg, query, gender, locale), item features (\eg, popularity), and user-item interaction features (\eg, was it seen by the user). The plain input data will repeat user and item features many times since the training data store all user item interaction pairs, which results in hundreds of terabytes of disk space occupation, more data transferring time and slow training speed. To solve this problem, we customized TensorFlow Dataset~\cite{tf_dataset} to assemble training examples from three separate files, a user feature file, an item feature file and an interaction file with query, user id and item id. The user and item feature files are first loaded into memory as feature lookup dictionaries, then the interaction file is iterated over the training steps with the user and item features appended. With this optimization, we successfully reduced the training data size to be $10\%$ of the original size.

\subsubsection{Scalable Distributed Training}
In the scenario of distributed training with parameter servers, one of the common bottlenecks is network bandwidth.
Most of mainframe network bandwidth in industry is 10G bits that are far from enough for large deep learning models. We observed that the off-the-shelf TensorFlow Estimator implementation is not optimized enough when handling embedding aggregation (\eg, sum of embeddings), thus the network bandwidth becomes a bottleneck quickly while adding a handful of workers. To further scale up the training speed, we improved the embedding aggregation operator in TensorFlow official implementation by moving the embedding aggregation operation inside parameter server, instead of in the workers. Thus, only one embedding is transferred between parameter server and worker for each embedding aggregation, instead of tens of them. Therefore, network bandwidth is significantly reduced, and the distributed training can be scaled up to five times more machines.

\subsection{Online Serving System}
\label{sec:serving_system}
The overview of {\dpsr} online serving system is shown in Figure~\ref{fig:serving}. The system consists of two novel parts that we would like to elaborate on, one TensorFlow Servable~\cite{tfserving} model, and a proxy for model sharding.

\subsubsection{One Servable Model}
The straightforward implementation of {\dpsr} can be composed of two separate parts, query embedding computation, and nearest neighbor lookup. 
Without careful design, one can simply build two separate online services for them. However, this is not the optimal system design in the sense of two points, a) it introduces complexity to manage the mapping between query embedding model and item embedding indexes, which could completely cause system failure if mapping mistake happens. b) it needs two network round trips to compute the nearest neighbors for a given query text. To overcome these issues, we take a more optimized approach by utilizing TensorFlow Servable~\cite{tfserving} framework, where we can unify the two parts into one model. As shown in Figure~\ref{fig:serving}, the two parts can be encapsulated into one Servable. The query embedding is sent directly from query embedding model to item embedding index, via computer memory, instead of via computer network.

\subsubsection{Model Sharding}
\begin{figure}[t]
    \centering
    \includegraphics[width=0.45\textwidth]{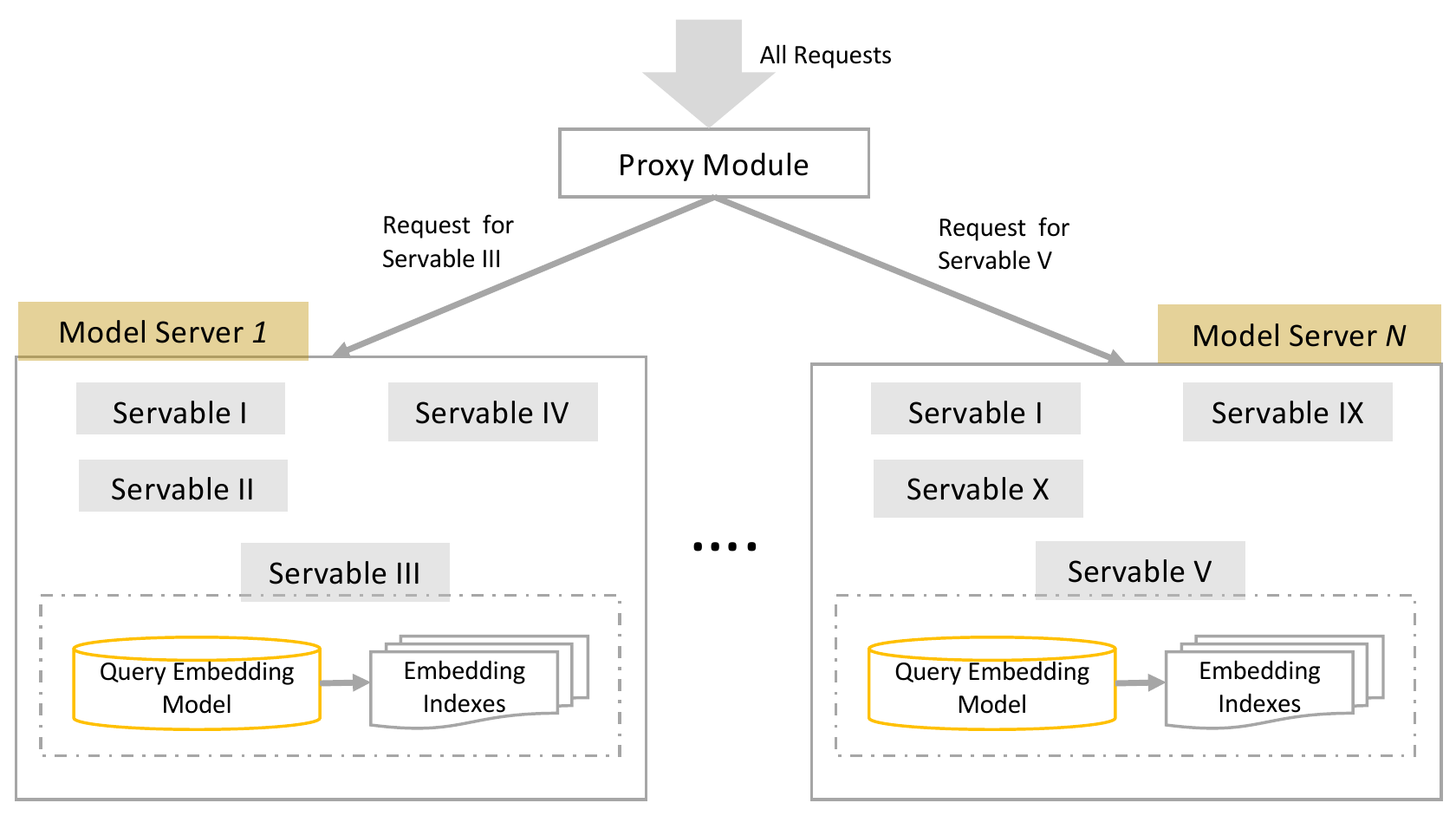}
    \vspace{-2mm}
    \caption{Online serving system for DPSR.}
    \label{fig:serving}
\end{figure}

The further scale up of the system needs to support hundreds of {\dpsr} models online at the same time, for different retrieval tasks, and for various model A/B experiments. However, one servable model consisting of one query embedding model and one item embedding index usually takes tens of Gigabytes of memory. Thus, It becomes infeasible to store all the models in one machine's memory, and we have to build a system to support serving hundreds of {\dpsr} models. We solve this problem by a proxy module, which plays the role of directing model prediction requests to one of the model servers that hold the corresponding model, as shown in Figure~\ref{fig:serving}. This infrastructure is not only designed for {\dpsr}, but as a general system for supporting all deep learning models at our search production.

% \subsubsection{Batching}
% To further increase serving capacity, we develop a batching techniques when the query per second (QPS) is high. Basically, instead of performing  model prediction immediately after receiving the request, we could buffer a few requests and send them together to the GPU for model prediction. By this, we are able to increase the QPS by about $33\%$, with slight trade-off of latency for about $2$ milliseconds.

\section{Experiments}
\label{sec:experiments}
In this section, we first visualize the embedding results leveraging t-SNE in Section~\ref{sec:ablation-study}, so we can get the intuition of how the model works.
Then we report offline evaluations by comparing with different methods in Section~\ref{sec:offline-evaluation}.
Next, we report online A/B test results in our search production, one of largest e-commerce search engines in the world, in Section~\ref{sec:online-evaluation}. 
Furthermore, we also report the offline indexing and online serving time of our {\dpsr} system in Section~\ref{sec:efficiency}, to demonstrate its efficiency, which is crucial in the industrial world.

Our production {\dpsr} model is trained on a data set of $60$ days user click logs, which contains $5.6$ billion sessions. We conducted distributed training in a cluster of five $48$-cores machines, with a total of $40$ workers and $5$ parameters servers launched. We used margin parameter $\delta=0.1$, AdaGrad~\cite{duchi2011adaptive} optimizer with learning rate $0.01$, batch size $b=64$, embedding dimension $d=64$. The training converges in about $400$ million steps for about $55$ hours.

\subsection{Embedding Visualization and Analysis}
\label{sec:ablation-study}

\iffalse
\begin{table}[tb]
    \centering
    \caption{Comparison between baseline methods with the proposed {\dpsr}.}
    \begin{tabular}{c|ccccc}
    \hline
        & Top-$1$   & Top-$10$ & MRR & AUC & Time  \\
    \hline
    BM2.5 & $0.718$ & $0.947$ & $0.802$ & $0.661$ & $61$ s \\
    BM2.5-u\&b & $0.721$ & $0.948$ & $0.802$ & $0.661$ & $157$ s  \\
    BERT & $0.031$ & $0.117$ & $0.065$ & $0.714$ & $146$ ms \\
    DSSM & $0.002$  &  $0.016$ & $0.01$ & $0.524$ & $20$ ms\\
    DPSR & $0.860$ & $0.993$ & $0.915$ & $0.757$ & $20$ ms\\
    DPSR-p & $0.845$ & $0.982$ & $$ & $$ & $20$ ms\\
    DPSR-h & $0.878$ & $0.998$ & $$ & $$ & $20$ ms\\
    \hline
    \end{tabular}
    \label{tab:batch_negative}
\end{table}
\fi

\begin{figure*}[t]
\centering
\includegraphics[width=0.70\textwidth]{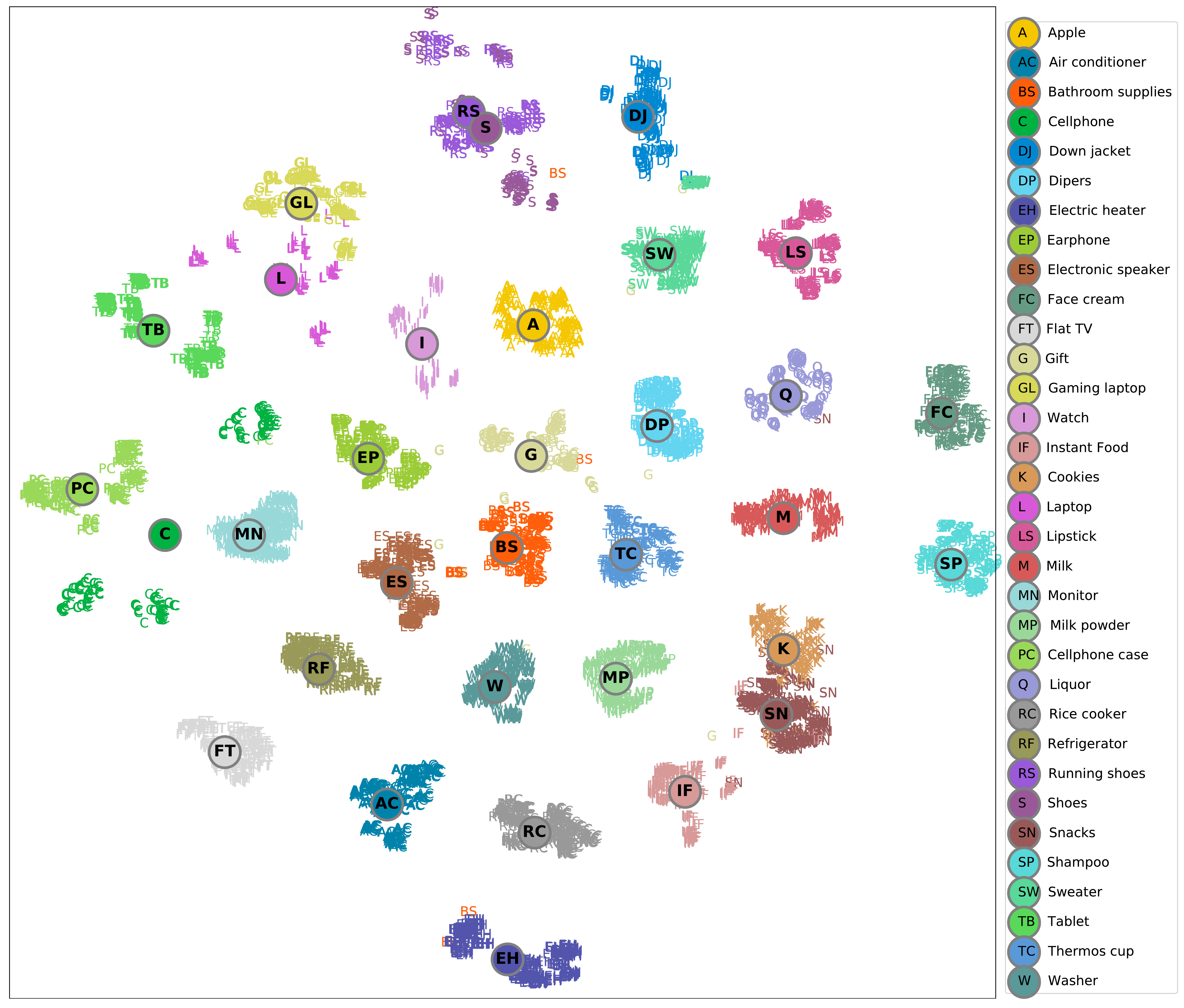}
\vspace{-3mm}
\caption{t-SNE visualization of item embeddings from 33 most popular categories.}

\label{fig:topology}
\end{figure*}

\subsubsection{Embedding Topology}
To have an intuition of how our embedding retrieval model works, we illustrate the $2$-D t-SNE coordinates for frequent items chosen from the most popular $33$ categories in our platform. As shown in Figure~\ref{fig:topology}, we can see that the item embeddings are structured in a very explicit and intuitive way. Basically, we can see that the electronics related categories, \eg, phones, laptops, tablets, earphones, monitors are well placed on the left side of the figure. The appliance related categories, \eg, refrigerator, flat TV, air conditioner, washer and so on are placed on the lower left side.
The food related categories, \eg, snacks, instant food, cookies, milk powder, are placed on the lower right part. The cleaning and beauty related categories, \eg, face cream and shampoo, are placed on the right part. 
The clothes related categories, \eg, shoes, running shoes, sweaters and down jackets, are placed on the upper right part. Overall, this reasonable and intuitive embedding topology reflects that the proposed model well learns the item semantics, which in turn enables query embeddings to retrieve relevant items.

\subsubsection{Multi-Head Disambiguation}
\begin{figure*}[h]
    \centering
    \begin{subfigure}[b]{0.22\textwidth}
    \centering
    \includegraphics[height=45.5mm]{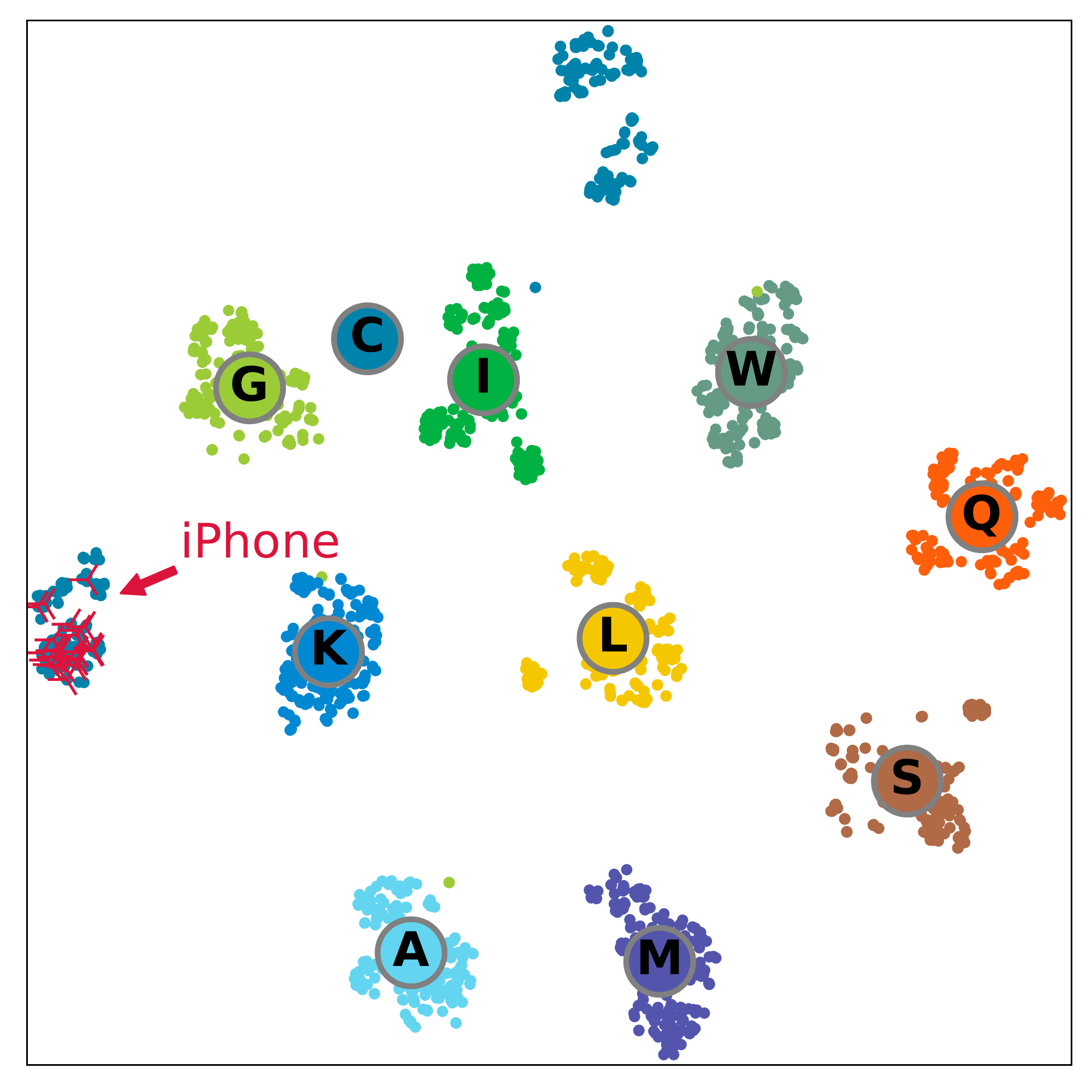}
    \caption{One head for query ``apple''.}
    \label{fig:1_head}
    \end{subfigure}
    \hfill
    \begin{subfigure}[b]{0.33\textwidth}
    \centering
    \includegraphics[height=47mm]{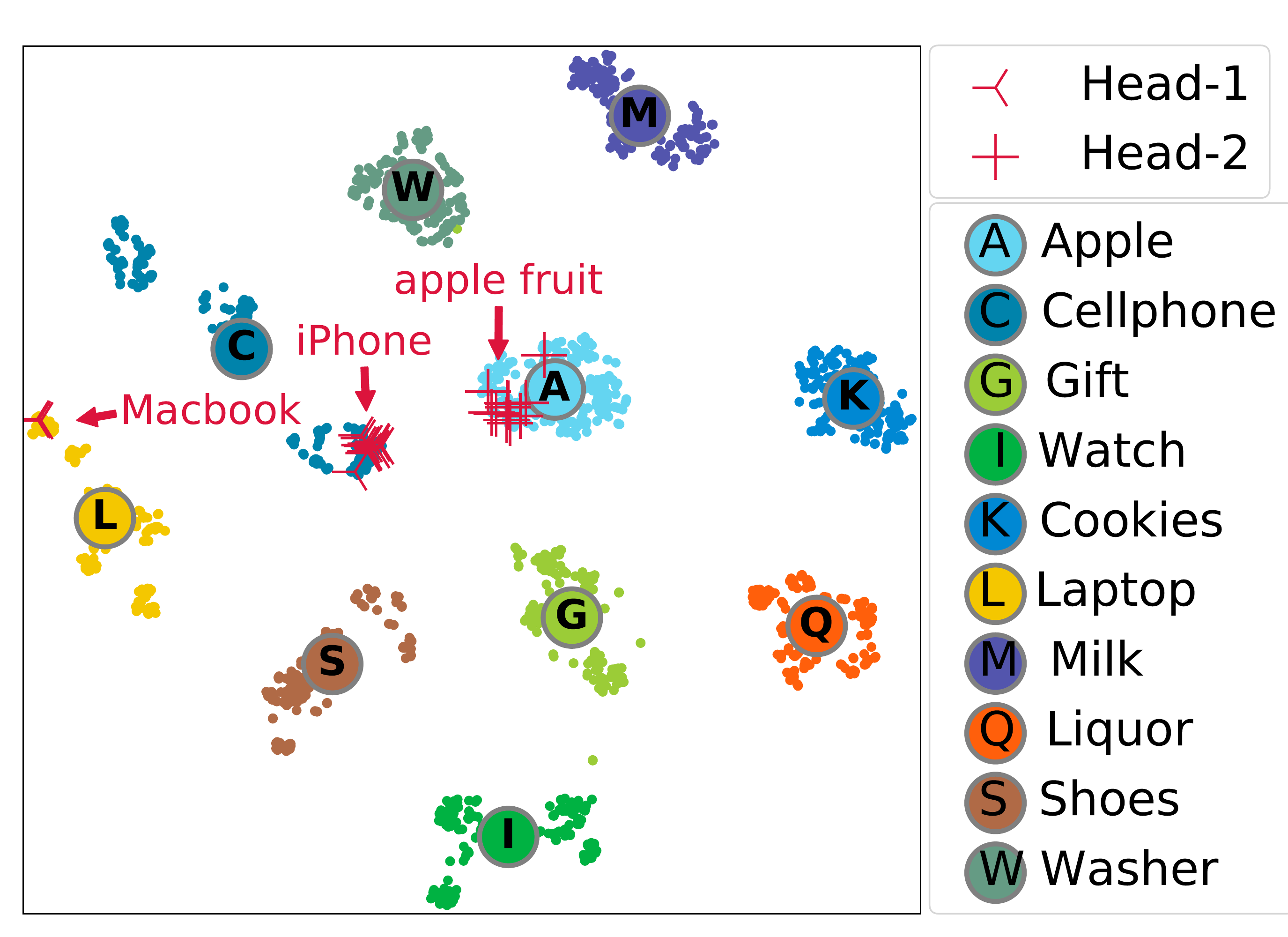}
    \caption{Two heads for query ``apple''. }
    \label{fig:2_head}
    \end{subfigure}
    \hfill
    \begin{subfigure}[b]{0.35\textwidth}
    \centering
    \includegraphics[height=47mm]{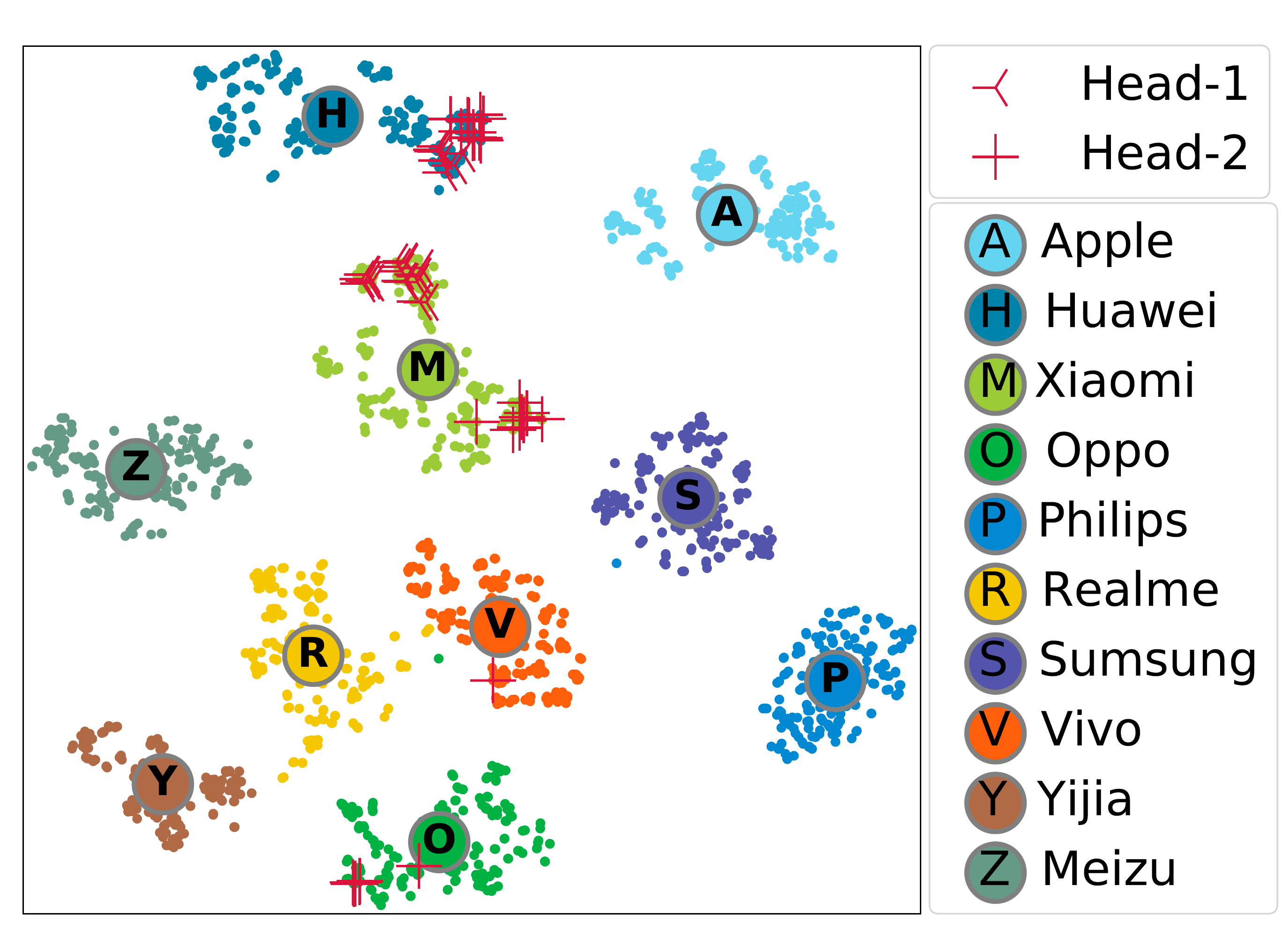}
    \caption{Two heads for query ``cellphone''.}
    \label{fig:2_head_cellphone}
    \end{subfigure}
    \caption{t-SNE visualizations of retrieval results for polysemous queries.}
    \label{fig:all_emb_pca}
\end{figure*}

% Figure~\ref{fig:2_head} and \ref{fig:2_head_cellphone} also show two query head embeddings for ambiguous query ``apple'', and ``cellphone''. 
In Figure~\ref{fig:2_head}, we also compute the $2$-D t-SNE coordinates for frequent items chosen from $10$ commodity categories to illustrate the effect of having multi-heads in query tower. We use two polysemous queries as an example here, ``apple'' and ``cellphone'', which are also within the top-10 queries in our platform.
We can see that the two heads for the query ``apple'' separately retrieve iPhone/Macbook and apple fruit. In Figure~\ref{fig:2_head_cellphone}, we can see that the two heads for the query ``cellphone'' retrieve the two most popular brands, Huawei and Xiaomi, separately. The illustration shows that different heads are able to focus on different possible user intentions.
In contrast, the single head model in Figure~\ref{fig:1_head} does not cluster well for cellphone category, where the iPhones are forming another cluster far away from other cellphones, potentially due to the ambiguity of the very top query ``apple''.

\subsubsection{Semantic Matching}
For better understanding of how our proposed model performs, we show a few good cases from our retrieval production in Table~\ref{tab:good_cases}. We can observe that {\dpsr} is surprisingly capable of bridging queries and relevant items by learning the semantic meaning of some words, such as big kid to 3-6 years old, free-style swimming equipment to hand paddle, and grandpa to senior. Also, {\dpsr} is able to correct typos in the query, such as v bag to LV bag, and ovivo cellphone to vivo cellphone, partially because we leverage English letter trigrams in the token vocabulary. We also observed similar typo corrections for Chinese characters, which are mainly learned from user clicks and $n$-gram embeddings.

\begin{CJK*}{UTF8}{gbsn}

\begin{table*}[t]
    \centering
    \caption{Good cases from {\DPSR} system in production.}
    \vspace{-2mm}
    \label{tab:good_cases}
    \begin{tabular}{c|cc}
    \hline
    query & retrieved item \\
    \hline
     奶粉$\;$大童\quad(milk powder big kid)   &  美赞臣$\;$安儿健A+$\;$4段 \quad(Enfamil A+ level-4 for 3 to 6 years old) \\
     \hline
     碧倩套装 \quad(``Clinique typo'' set)   &  倩碧(CLINIQUE)经典三部曲套装 \quad(Clinique classic trilogy set) \\
     \hline
    官网v女包 \quad(authentic v women bag)  & 路易威登LV女包 \quad (Louis Vuitton LV women bag) \\
     \hline
     ovivo手机 \quad (ovivo cellphone)  & vivo Z1\quad (vivo Z1 phone) \\
     \hline
     学习自由泳器材\quad (learn free-style swimming equipment)   & 英发/yingfa 划臂 \quad(yinfa hand paddle)\\
    \hline
    \end{tabular}
\end{table*}
\end{CJK*}

\subsection{Offline Evaluations}
\label{sec:offline-evaluation}
\subsubsection{Metrics}
We use the following offline metrics to evaluate the retrieval methods.

{Top-$k$} is defined as the probability that a relevant item is ranked within the top $k$ retrieved results among $N$ (we used $1,024$) random items for a given query. This top-$k$ value is empirically estimated by averaging $200,000$ random queries. A higher top-$k$ indicates a better retrieval quality, \ie, hit rate.

%{MRR} is multiplicative inverse of the rank of the relevant item. Similar as top-$k$, the higher MRR, the better performance.

{AUC} is computed in a separate data set with human labeled relevance for query item pairs. The labels can be categorized into relevant and non-relevant ones, and then the embedding inner products or any relevancy scores (BM2.5) can be treated as prediction scores. A higher AUC here indicates a better retrieval relevancy.

{Time} is the total retrieval time on a 48-core CPU machine from a query text to $1,000$ most relevant items out of a set of $15$ million items. This metric value decides whether a method is possible to apply to industry-level retrieval system or not. Typically, the cutoff is $50$ milliseconds, but preferably $20$ milliseconds.

\subsubsection{Baseline Methods}
We compared {\dpsr} with BM2.5 and DSSM as baselines.
{BM2.5} is a classical information retrieval method based on keywords matching using inverted index, and it uses heuristics to score documents based on term frequency and inverted document frequency. We compare with two versions of BM2.5, with only unigrams, and with both unigrams and bigrams (denoted as BM2.5-u\&b).
{DSSM} is a classical deep learning model~\cite{Huang:2013:LDS:2541176.2505665} designed for ranking but not retrieval. We still would like to include the comparison to clarify the difference.

\subsubsection{Results}

In Table~\ref{tab:batch_negative}, we show the comparison results with the above baseline methods.
We can make the following observations from the results.

\begin{itemize}
\item BM2.5 as a classical method shows good retrieval quality, but it takes more than a minute to retrieve from $15$ million items, which means that it is too unrealistic to use it in online retrieval. 
\item DSSM that samples unclicked items as negative examples performs worst in top-$k$, MRR and AUC. This is mainly due to that DSSM is optimized for ranking tasks, which is a highly different task from retrieval. Therefore, we can conclude that only using unclicked items as negative examples does not work to train a retrieval model.
\item {\dpsr} refers to a vanilla version of our model without any user features. It has the highest AUC score among the baseline methods and other personalized {\dpsr} versions, which indicates that pure semantic {\dpsr} could achieve the highest retrieval relevance. % Also the retrieval time, $20$ milliseconds, is fast enough for a online retrieval system.
\item {\dpsr-p} refers to a basic personalized version of our model, with additional user profile features, like purchase power, gender and so on. The result shows that those profile features help improve the retrieval quality metrics (Top-k) over the vanilla version, with a slight tradeoff of relevancy.
\item {\dpsr-h} refers to a full personalized version of our model, with both user profile and user history events. It has the best retrieval quality metrics (Top-k) over all models, which demonstrates that plenty of signals can be squeezed from the user history events. Note that the personalized model improves the retrieval quality metrics with a tradeoff of relevance metrics (AUC), which is also reasonable, since the retrieval quality consisting of more factors besides relevancy, such as item popularity, personalization and so on.
\end{itemize}

\begin{table}[tb]
    \centering
    \caption{Comparison between different methods.}
    \vspace{-2mm}
    \label{tab:batch_negative}
    \begin{tabular}{c|ccccc}
    \hline
        & Top-$1$   & Top-$10$  & AUC & Time  \\
    \hline
    BM2.5 & $0.718$ & $0.947$ & $0.661$ & $61$ s \\
    BM2.5-u\&b & $0.721$ & $0.948$  & $0.661$ & $157$ s  \\
    DSSM & $0.002$  &  $0.016$  & $0.524$ & $20$ ms\\
    DPSR & $0.839$ & $0.979$ & $\mathbf{0.696}$ & $20$ ms\\
    DPSR-p & $0.868$ & $0.984$  & $0.692$& $20$ ms\\
    DPSR-h & $\mathbf{0.889}$ & $\mathbf{0.998}$  & $0.685$ & $20$ ms\\
    \hline
    \end{tabular}
\end{table}

\begin{figure}[t]
    \centering
    \includegraphics[width=0.35\textwidth]{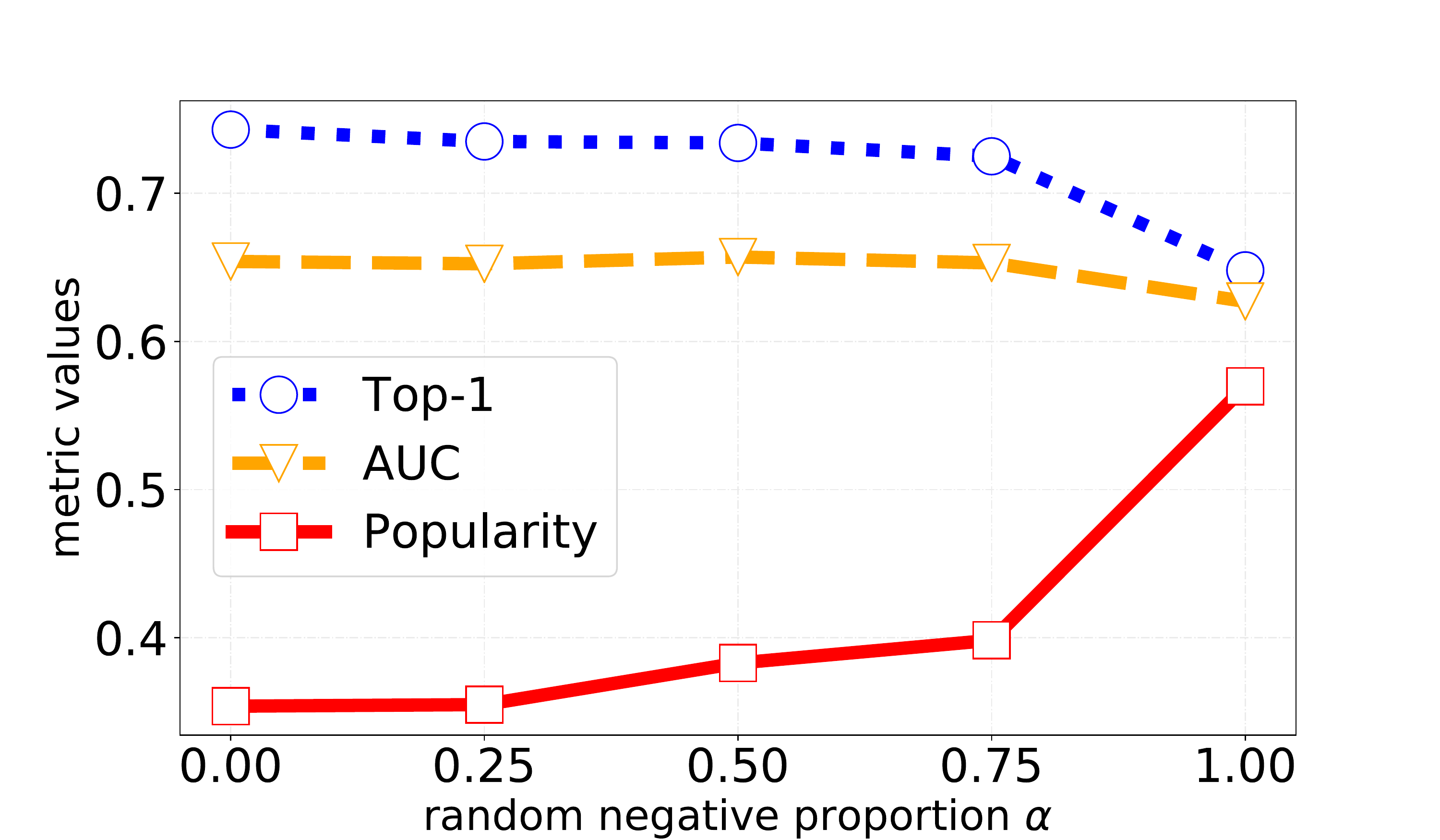}
    \vspace{-2mm}
    \caption{Effect with different mixing ratio of negatives.}
    \label{fig:trend_alpha}
\end{figure}

Moreover, Figure~\ref{fig:trend_alpha} illustrates that the mixing ratio $\alpha$ of random negatives and batch negatives (see Section~\ref{sec:mixing_ratio}) affects the retrieved item popularity.
Basically, we can observe that the more random negatives we have in the negative sampling, the more popular items are retrieved. But too many random negatives, \eg, $\alpha=1.0$, will hurt the retrieved item's relevancy. Thus, we can treat the parameter of $\alpha$ as a tradeoff between retrieval popularity and relevancy. In practice, we also found a proper choice of $\alpha=0.5$ or $\alpha=0.75$ would help online metrics significantly.

\subsection{Online A/B Tests}
\label{sec:online-evaluation}
{\dpsr} is designed as a key component in our search system to improve the overall user experience. Thus, we would like to focus on the overall improvement of a search system using {\dpsr} as an additional retrieval method.

In the control setup (baseline), it includes all the candidates available in our current production system, which are retrieved by inverted-index based methods with query rewritten enabled. In the variation experiment setup ({\dpsr}), it retrieves at most $1,000$ candidates from our {\DPSR} system in addition to those in the baseline. For both settings, all the candidates will go through the same ranking component and business logic. The ranking component applies a state-of-the-art learning-to-rank method similar to methods mentioned in~\cite{Liu:2017:CRO:3097983.3098011}. Here, we emphasize that our production system is a strong baseline to be compared with, as it has been tuned by hundreds of engineers and scientists for years, and has applied state-of-the-art query rewriting and document processing methods to optimize candidate generation.

We first conducted human evaluation for the relevance of retrieved items. Specifically, we ask human evaluators to label the relevance of results from the baseline system and {\dpsr} for the same set of $500$ long tail queries. The labeled results are categorized into $3$ buckets, bad, fair and perfect. Table~\ref{tab:relevancy_labeling} shows that the proposed method improve search relevancy by reducing around $6\%$ bad cases. It proves that the deep retrieval system is especially effective in handling ``difficult'' or ``unsatisfied'' queries, which often require semantic matching.

\begin{table}[tb]
    \centering
    \caption{Relevancy metrics by human labeling of 500 long tail queries. {\dpsr} reduces bad cases significantly.}
    \vspace{-2mm}
    \label{tab:relevancy_labeling}
    \begin{tabular}{c|c c c}
    \hline
        & bad   & fair  & perfect \\
    \hline
    Baseline    & $17.86\%$   & $26.04\%$   & $56.10\%$ \\
    {\DPSR}    & $13.70\%$   & $33.28\%$ & $53.01\%$ \\
    \hline
    \end{tabular}
\end{table}

We then conducted live experiments over $10\%$ of the entire site traffic during a period of two weeks using a standard A/B testing configuration. To protect confidential business information, only relative improvements are reported. Table~\ref{tab:abtest} shows that the proposed DPSR retrieval improves the production system for all core business metrics in e-commerce search, including user conversation rate (UCVR), and gross merchandise value (GMV), as well as query rewrite rate (QRR), which is believed to be a good indicator of search satisfaction. We can also observe that the 2-heads version of query tower, and personalized version (denoted as 1-head-p13n) both improve the vanilla version of 1-head query tower without any user features. Especially, we observe that the improvements mainly come from long tail queries, which are normally hard for traditional search engines.

%This coincides with that query rewrite rate (QRR) has been greatly reduced, which means our method is very effective to the queries that are likely to require additional users efforts. Those queries are generally difficult to handle, and are likely to be long tail queries.

\begin{table}[tb]
    \caption{{\DPSR} Online A/B test improvements.}
    \vspace{-2mm}
    \centering
    \begin{tabular}{c|r r r r}
    \hline
            & UCVR  & GMV  & QRR \\
    \hline
    1-head  & $+1.13\%$ & $+1.78\%$ & $-4.44\%$ \\
    2-head  & $+1.34\%$ & $+2.13\%$ & $-4.13\%$ \\
    1-head-p13n & $+1.29\%$ & $+2.19\%$ & $-4.29\%$ \\
    \hline
    2-head on long tail query  & $+10.03\%$ & $+7.50\%$ & $-9.99\%$ \\
    \hline 
    \end{tabular}
    \label{tab:abtest}
\end{table}

% For better understanding of how our proposed model performs, we show a few good cases from our {\DPSR} production in Table~\ref{tab:good_cases}. We can observe that {\DPSR} is surprisingly capable of bridging queries and relevant items by learning the semantic meaning of some words, such as big kid to 3-6 years old, free-style swimming equipment to hand paddle, and grandpa to senior. Also, {\DPSR} is able to correct typos in the query, such as v bag to LV bag, and ovivo cellphone to vivo cellphone, partially because we leverage English letter trigrams in the token vocabulary. We also observed similar typo corrections for Chinese characters, which are mainly learned from user clicks and $n$-gram embeddings.

% Furthermore, we show in Table~\ref{tab:sale_boosted} that incorporating sales information in our model can achieve better CVR and GMV metrics than the one with only click information. 

%we randomly selected a set of queries (\i.e., 500), and, for each query, %we picked results from each system at the similar positions. Human labeler evaluates each pair by a 7-level scoring system, which are then categorized into 3 buckets, namely bad, passable and good

\subsection{Efficiency}
\label{sec:efficiency}
In Table~\ref{tab:gpu}, we show the efficiency of our offline index building and online nearest neighbor search excluding the query embedding computation. 
We report the time consumed for indexing and searching $15$ million items with NVIDIA Tesla P40 GPU and Intel $64$-core CPU. It shows that {\dpsr} can retrieve candidates within $10$ms on CPU, and can benefit from GPUs with $85\%$ reduction on indexing time consumption, $92\%$ reduction on search latency and $14$ times more QPS (query per second) throughput.

In Table~\ref{tab:overall_latency}, we report the overall model serving performance with the same CPU and GPU machines as above. The overall latency from query text to $1,000$ nearest neighbors can be done within $15$ to $20$ milliseconds for GPU or CPU machines, which is even comparable to the retrieval from standard inverted index. %Note that the results in Table~\ref{tab:gpu} and Table~\ref{tab:overall_latency} are obtained under different QPS stress testing conditions.

\begin{table}[tb]
    \centering
    \caption{Latency for index building and search.}
    \vspace{-2mm}
    \label{tab:gpu}
    \begin{tabular}{c|c c c}
    \hline
        &  index building (sec.)  &  search (ms) & QPS\\
    \hline
    CPU   & $3453$   & $9.92$ & $100$\\
    GPU  & $499$  & $0.74$ & $1422$ \\   
    \hline
    \end{tabular}
    
\end{table}

\begin{table}[tb]
    \centering
    \caption{Overall serving performance.}
    \vspace{-2mm}
    \label{tab:overall_latency}
    \begin{tabular}{c|c c c c}
    \hline
        &  QPS & latency (ms) & CPU usage & GPU usage\\
    \hline
    CPU   & ~$4,000$   & $20$ & $>50\%$ & $0.0\%$\\
    GPU  & $5,800$ & $15$ & $>50\%$ & $25\%$ \\   
    \hline
    \end{tabular}
\end{table}

% We report latency of the deep retrieval system in Table~\ref{tab:latency}. The latency is measured by stress tests on a single machine with $64$-cores CPU and $256$ GB at different query per seconds (QPS) level. The results show that our system can handle requests efficiently and does not add any significant latency on the existing search system.  

\section{Conclusion}
In this paper, we have discussed how we build a deep personalized and semantic retrieval system in an industry scale e-commerce search engine. Specifically, we 1) shared our design of a deep retrieval system,  which takes all the production requirements into consideration, 2) presented a novel deep learning model that is tailored for the personalized and semantic retrieval problems, and 3) demonstrated that the proposed approach can effectively find semantically relevant items, especially for long tail queries, which is an ideal complementary candidate generation approach to the traditional inverted index based approach. We have successfully deployed {\dpsr} into JD.com's search production since early 2019, and we believe our proposed system can be easily extended from e-commerce search to other search scenarios. 

\section{Acknowledgement}
We deeply appreciate Chao Sun, Jintao Tang, Wei He, Tengfei Guan, Wenbin Zhu, Dejun Qiu, Qi Zhu, Hongwei Shen, Wei Wei and Youke Li for their engineering support to build key components of the infrastructure, and Chen Zheng, Rui Li and Eric Zhao for their help at the early stage of this project. We thank the anonymous reviewers for their valuable suggestions.

\bibliographystyle{ACM-Reference-Format}
\balance

\bibliography{bibliography}

\end{document}